\documentclass[letter]{aa}  
\usepackage{graphicx}
\usepackage{natbib}
\usepackage{txfonts}

\begin{document}
   \title{A brown dwarf companion to the intermediate-mass star HR\,6037\thanks{Based on observations 
collected at the Paranal Observatory under programs 272.D-5068(A), 77.D-0147(A), and 285.C-5008(A)}}
   \author{N. Hu\'elamo\inst{1}
          \and
          D. E. A. N\"urnberger\inst{2} 
          \and
          V. D. Ivanov\inst{2}
          \and
          G. Chauvin\inst{3}
          \and
          G. Carraro\inst{2,4}
          \and 
          M. F. Sterzik\inst{2}   
          \and 
          C. H. F. Melo\inst{2}
          \and
          \mbox{M. Bonnefoy}\inst{3} 
          \and
          M. Hartung\inst{5}
          \and
          X. Haubois\inst{6}
         \and
          C. Foellmi\inst{3}
          }
\offprints{N. Hu\'elamo}
\institute{Centro de Astrobiolog\'{\i}a (CSIC-INTA); LAEFF, P.O. Box 78, E-28691 Villanueva de la Ca\~nada, Madrid, Spain\\
         \email{nhuelamo@cab.inta-csic.es}
         \and 
         European Southern Observatory, Alonso de Cordova 3107, 
         Casilla 19, Santiago, Chile
         \and 
         Laboratoire d'Astrophysique, Observatoire de Grenoble, BP 53, 
         38041 Grenoble, Cedex 9, France
         \and
         Universit\'a di Padova, Vicolo Osservatorio 3, I-35122, Padova Italy
         \and
         Gemini Observatory, Southern Operations Center, c/o AURA, Casilla 603, La Serena, Chile         
         \and
         Instituto de Astronom\'{\i}a, Geof\'{\i}sica e Ciencias Atmosf\'ericas, Rua do Mat\~ao, 1226- Cidade Universitaria, S\~ao Paulo, Brasil}
   \date{Received; accepted}

 
  \abstract
   {The frequency of brown dwarf and planetary-mass companions to intermediate-mass stars is still unknown.
   Imaging and radial velocity surveys
   have revealed a small number of substellar companions to 
   these stars.}
   {In the course of an imaging survey we have detected a visual companion to the intermediate-mass 
   star HR\,6037. In this letter we confirm it as a co-moving substellar object.
 }
   {We present two epoch adaptive optics observations of HR\,6037, an A6-type 
   star with a companion
   candidate at 6\farcs67 and position angle of 294 degrees. We also analyze
   near-infrared spectroscopy of the companion.}
   {Two epoch observations of HR\,6037 have allowed us to confirm 
   HR\,6037\,B as a co-moving companion.   Its $J$ and $H$ band spectra suggest the object has an spectral type of M9, with a surface
   gravity intermediate between a 10\,Myr dwarf and a field dwarf of the same spectral type.
   The comparison of its $K_s$-band photometry with evolutionary tracks allows us to derive a mass, effective 
   temperature, and surface gravity of 62$\pm20$\,M$_{Jup}$, $T_{\rm{eff}} = 2330\pm200$\,K, and log\,$g= 5.1\pm0.2$, respectively.
   The small mass ratio of the binary, $\sim$0.03, and its long orbital period, $\sim${\bf 5000}\,yr,  makes HR\,6037 a
   rare and uncommon binary system.
    }
   {}


   \maketitle
%

\section{Introduction}



The frequency of brown dwarfs (BDs) and planetary-mass companions around intermediate-mass main sequence (MS) stars is  uncertain.  BDs  can be formed by several  mechanisms
\citep[e.g.][]{Padoan2004,Stamatellos2009}, but the expected substellar fractions for B-F type primaries 
are uncertain. In the case of giant planets formed in the disks of primary stars, some works predict a higher frequency around AB-type stars than in solar-type stars  \citep[e.g.][]{Kennedy2008}.   However, \citet{Kornet2006} show  an opposite result since they conclude that the percentage of stars with giant planets decreases with increasing stellar masses from 0.5 to 4\,M$_{\odot}$.

To shed light on this issue,  different observational programs have 
been focused on deriving the frequency of BDs and planetary-mass objects around intermediate-mass stars.
As a result, planetary mass companions have been recently detected around three A-type stars through 
adaptive optics (AO) assisted
observations \citep[][]{Marois2008, Kalas2008, Lagrange2010}. 
Radial velocity (RV) studies, which are sensitive to short period companions, have also
reported the presence of substellar objects
around several A-F type MS stars \citep{Galland2005,Galland2006,Guenther2009}, with minimum
masses ($M. sin i$ ) in the planetary mass regime, that is,  they could also be BDs.
Transit programs have also detected planetary mass companions around several F-type stars 
\citep[e.g.][]{Bakos2007,JohnsKrull2008,Joshi2009,Hellier2009}, and one around an A5 star  
\citep{Christian2006,Cameron2010}. 

In the case of BDs,  direct imaging surveys have allowed to study the fraction of wide substellar companions
around intermediate-mass stars. As an example, \citet{Kouwenhoven2005,Kouwenhoven2007} studied the late-B and A-type star population from the  Sco OB2 association.
Although sensitive to substellar companions, they reported the detection of only two
BDs companion candidates.  
They concluded that the dearth of BD companions to
intermediate-mass stars is consistent with the extrapolation of the stellar companion mass distribution into 
the BD regime (assuming they formed like stars). 
Recently, \citet{Ehrenreich2010} conducted an AO survey to detect substellar companions
in wide orbits around a volume-limited sample of 38 A- and F-type field stars previously observed with RV techniques.
They did not report any new BD companion.
In fact, up to now there is only one BD companion to an intermediate-mass star,
HR7329\,B, confirmed by direct imaging and near-IR spectroscopy \citep[][]{Lowrance2000,Guenther2001}.
RV studies have also detected BD companions to  A-F type MS stars
\citep{Galland2006,Hartmann2010}, and
transit observations have reported the presence of a BD  around an F-type star,  CoRoT-3\,b \citep{Deleuil2008}.

As it follows, and despite the efforts, the occurrence of
planetary-mass objects and BDs around intermediate-mass stars is still
unknown and deserves additional observations.

In 2004, we started a project aiming at deriving the binary fraction
and properties among a large volume-limited sample of
intermediate-mass stars in the field \citep[hereafter, Multi-NETS
  Project,][]{Ivanov2006}. Thanks to the use of deep AO near infrared
imaging with Naos-Conica \citep[NACO][]{Lenzen2003} at the Very Large
Telescope (VLT), we have been able to extend our study to substellar
companions that are important to understand the formation of low mass
ratio binaries. In the course of our survey, we detected a faint,
visual companion to the star HR\,6037.  In this letter, we report the
discovery and the co-moving confirmation of the companion to HR\,6037
based on NACO astrometric observations obtained at two different
epochs. We also present ISAAC near-infrared spectroscopic data
that confirms, together with the photometry, that
this companion is likely to be a new and rare substellar companion to
the intermediate-mass star HR\,6037.

\section{HR\,6037 stellar properties}

   
\begin{figure}
   \centering
   \includegraphics[width=0.8\columnwidth]{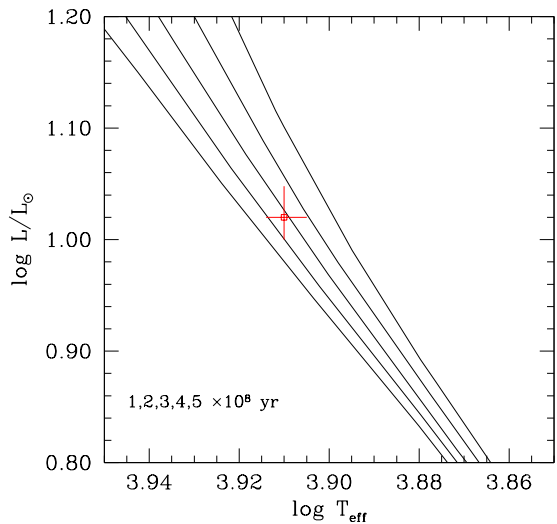}
     \caption{HR Diagram with the location of HR\,6037. 
      We have overplotted the \citet{Marigo2008} 
       isochrones for five different ages between 100\,Myr and 500 Myr (from the left to the right). We estimate an age of
       $\sim$300$\pm$100 Myr for the primary. }
              \label{hrdiag}%
    \end{figure}

 HR\,6037
is a main sequence A-type star
classified as 'variable' by \citet{Samus2009} although its type of variability is uncertain.
Its proper motion and parallax, according to Hipparcos \citep{Perryman1997}, are $\mu_{\alpha}$=-44.74$\pm$0.56 and  $\mu_{\delta}$=-84.65$\pm$0.43\,mas/yr, and  18.13$\pm$0.69\,mas, respectively. The latter value 
translates into a distance of 55$\pm$2\,pc.

We have derived the physical properties of HR\,6037 by analyzing high resolution (R=80.000) optical
archival spectroscopy obtained with VLT/UVES  (Program ID 266.D-5655(A)). 
The spectrum was obtained integrating a total of 70 seconds, and was centered at 5800\AA.

The primary is an A6V, as derived from spectral synthesis using SYNTHE
\citep{Kuruc1993}.  Our analysis also yields [Fe/H]=0.00$\pm$0.05,
namely solar metallicity, and provides values for the effective
  temperature and surface gravity (see Table~\ref{properties}).
Using the Hipparcos parallax we derive $M_{V}=2.24$, or
$log(\frac{L}{L_{\odot}} $)=1.02$^{0.038}_{0.02}$, assuming no
extinction.  Stromgren H$_{\beta}=2.884$ and b-y$=0.069$ photometry
\citep{HauckMermilliod1998} confirms HR\,6037\,A as an A6 dwarf with
temperature and gravity in agreement with values derived from the UVES
spectrum.

We have plotted the object on a Hertzsrpung-Russel diagram (see Fig.~\ref{hrdiag}) and compared it with the Padova
evolutionary tracks for solar metallicity \citep{Marigo2008}. We have used isochrones from 100 to
500\,Myr, which is a typical age range for a star of this spectral type. 
We estimate an age of 300$\pm$100\,Myr and a mass of 1.8$\pm$0.2\,M$_\odot$.

Finally, we note that the object was included in a RV survey 
to look for very close BDs and planetary-mass
companions, showing no significant RV variation \citep{Lagrange2009}.

\section{Observations and data reduction}

\subsection{NACO deep imaging}

HR\,6037 was observed in service mode with NACO, the adaptive optics
facility at the VLT on June 30, 2004 and June 9, 2006. We used the
visible wavefront sensor with the primary as a reference star. We
observed in the Ks-band filter with the S27 objective 
(field of view of $27\arcsec\times27$\arcsec)
in 'Autojitter Mode', dithering within a box of 12\arcsec~width.
The total on-source exposure time was $\sim$13 minutes. 
The average coherence time and optical seeing 
were 1.5\,ms and 1\farcs0, 
and 1.2\,ms and 1\farcs7, during the first and second epoch, respectively.

\begin{figure}[t]
   \centering
   \includegraphics[width=0.7\columnwidth]{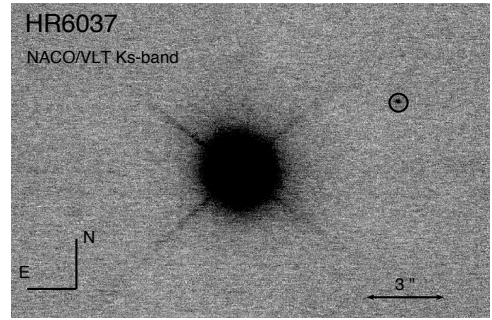}
   \caption{NACO/VLT image of HR\,6037. The co-moving companion is 
encircled.}
              \label{nacoima}%
    \end{figure}

The data were reduced using {\em Eclipse} \citep{Devillard1997}, and following the
standard procedure: dark subtraction, flat-field division, sky
subtraction, alignment and stacking. The final image from 2006 is displayed in
Fig.~\ref{nacoima}. Apart from the bright primary, we detect a visual companion at 
a projected separation of  $\sim$6\farcs66  and position angle of $\sim$294 degrees.

\begin{table}[t]
\caption{Physical properties of  HR\,6037\,A\&B derived in this work}
 \label{stellar}
\advance\tabcolsep by -2.8pt
\begin{tabular}{cclllc}
          \noalign{\smallskip} 
          \hline 
          \noalign{\smallskip}
 Name   & Sp. Type  & $T_{\rm eff}$ & log\,$g_{s}$ & Mass & [Fe/H] \\ 
 & & [K] & [cm/s$^2$] & [$M_{\odot}$]& \\
          \noalign{\smallskip} 
          \hline 
          \noalign{\smallskip}       

 HR6037A &    A6 &  8120$\pm$100  & 4.2$\pm$0.1 & 1.8$\pm$0.2 & 0.00$\pm$0.05 \\ 
 HR6037B &    M9 &  2330$\pm$200 & 5.1$\pm$0.2 & 0.06$\pm$0.02 & -- \\
      
\hline\hline
\end{tabular}\label{properties}
\end{table}

In order to measure precise separations and position angles, we
derived the plate scale and orientation of the detector, CONICA, using
archival observations of the astrometric calibrator IDS\,21506-5133 \citep{vanDessel1993}
obtained on June 2004 and August 2006. The
values for the two campaigns are respectively $27.01\pm0.05$~mas/pix
and $27.02\pm0.05$~mas/pix for the plate scale, and $0.0\pm0.2$~deg and
$-0.1\pm0.2$~deg for the True North orientation. 
   
\subsection{ISAAC near-infrared spectroscopy}

Near-infrared spectra of the HR\,6037\,B were obtained in service
mode with ISAAC/VLT  \citep{Moorwood1998}  on 2010-06-06/07
in the $J$ and $H$ atmospheric windows, and on 2010-06-11/12 in the $K$
window, in the ``classical'' nodding-along-the-slit observing
strategy. We used the low-resolution mode and the 0.6\arcsec ~wide slit,
delivering a spectral resolution of R$\sim$800. We collected six exposures for
$J$ and $H$, and twelve for $K$ but one $J$ spectrum was discarded because of
a low signal. The total integration times were 1115, 2232, and
4464\,sec, respectively for $J$, $H$, and $K$. The seeing was better than
1\arcsec~during both nights. The sky was clear on 2010-06-06/07, and
thin clouds were present on 2010-06-11/12. B-type telluric standards were
observed back-to-back with the science targets with the same instrument
setup.
One of them showed a strong Bracket\,$\gamma$ emission line, which was fitted with a Gaussian and
subtracted from the spectrum before applying the telluric correction.

The data were reduced using IRAF\footnote{IRAF is distributed by the
  National Optical Astronomy Observatory, which is operated by the
  Association of Universities for Research in Astronomy (AURA) under
  cooperative agreement with the National Science Foundation.}  and
following standard steps: flat field division, sky emission removal by
subtracting images from corresponding nodding pairs, and extraction
and combination of the individual spectra into the final spectrum.
The wavelength calibration was performed using arcs. The telluric
absorption was removed by divided the target spectra by the telluric
standards, and multiplied them by the corresponding spectra from the
library of \citet{Pickles1998}. Some of the spectra from this library
are featureless models, so artificial emission lines remained in the
final product. To remove them, we went back to the telluric spectra
and subtracted Gaussian fits to their intrinsic stellar features --
mainly Hydrogen recombination lines. This, together with the
Bracket\,$\gamma$ emission mentioned above implies some uncertainty in
the spectral regions around the strong Hydrogen lines.

\section{Results}

\subsection{HR\,6037\,B, a co-moving companion}

\begin{figure}[t]
   \centering
   \includegraphics[width=0.9\columnwidth]{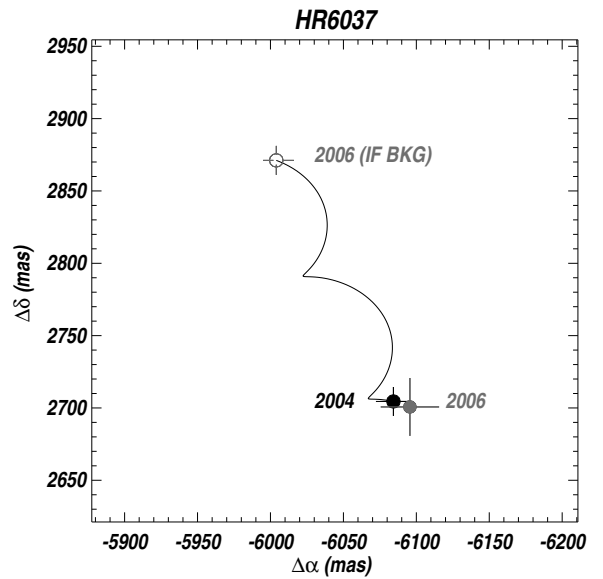}
   \caption{Analysis of the two NACO/VLT epochs of HR\,6037.  The axes show the difference
in right ascension and declination of the binary members in the two epochs. The solid black 
and grey circles represent the position of the
companion candidate in 2004 and 2006, respectively. The open grey circle
represents the expected position of HR\,6037\,B if it were a background object.
The data is consistent with HR\,6037\,B being a co-moving companion.}
              \label{ppm}%
    \end{figure}

Fig.~\ref{ppm} shows the difference in right ascension
(RA) and declination (DEC) of HR\,6037 and its companion candidate as
measured in 2004 and 2006. We have also overplotted the expected
difference in RA and DEC of a background object taking into account
the proper motion and parallax of the primary. As seen, the companion shows RA and
DEC differences consistent with a co-moving object.  In fact, 
the difference in separation
and position angle between the two epochs are consistent
with a bound companion within the errors (see Table~\ref{binary}).  


\subsection{Spectral characterization of HR\,6037\,B}

\begin{figure}[t]
   \includegraphics[width=0.5\textwidth]{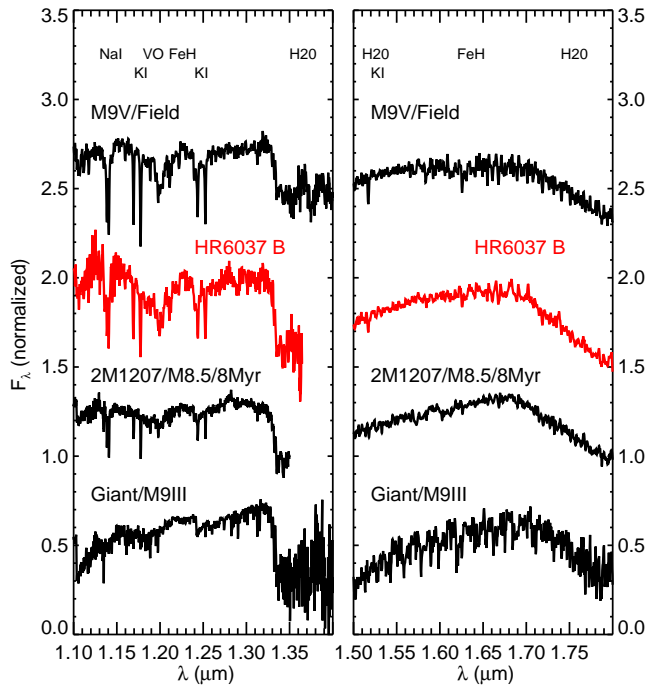}
   \caption{Comparison of the $J\&H$ band ISAAC spectra of
     HR\,6037\,B companion (red) to the template spectra of the young
     8\,Myr-old dwarf 2M1207\,A (M8.5) and the field dwarf (M9V). The
     spectrum of the very late type giant (IO Virginis) is also
     reported for comparison. All spectra have been normalized to 1.23~$\mu$m and 1.65~$\mu$m and offsetted.}
              \label{spectro}
    \end{figure}

The ISAAC $JHK$ spectra of HR\,6037\,B were first compared to
libraries of template spectra of field dwarfs
\citep{Cushing2005,Rayner2009} and moderately young dwarfs from Upper
Scorpius, TW Hydrae and $\beta$ Pictoris associations
\citep[][]{Allers2009, Rice2010}. The $K$-band spectrum of HR\,6037\,B
is much bluer than all M and early-L type dwarfs. This is probably due
to a problem of flux loss during the observation. Therefore, only the
$J$ and $H$-band spectra were considered for the spectral
classification based on the continuum comparison with libraries of
field and young dwarfs. The best matches are displayed in
Fig.~\ref{spectro}. The $J$ and $H$ continuum of HR\,6037\,B is well
reproduced by the spectrum of the young M8.5 dwarf 2M1207\,A from the
TW\,Hydrae association (8~Myr) and the M9V field dwarf
\citep{Rayner2009}, so we estimate an M9$\pm$1 spectral type.

Careful identification of the lines over the $JHK$ spectral range
shows the presence of broad molecular absorptions of H$_2$0 (longward
1.33 and 1.6$\mu$m), FeH (at 1.194, 1.222, 1.239, 1.583-1.591 and
1.625~$\mu$m) as well as CO overtones longward 2.29~$\mu$m all typical
of late-M dwarfs. There is also the possible presence of VO
absorptions from 1.17 to 1.20~$\mu$m. In the $J$-band, the atomic line
doublets of Na I and K I at 1.138, 1.169, 1.177, 1.243, 1.253~$\mu$m
are well detected.We also detect the K I atomic line at
1.517~$\mu$m. Their strengths are intermediate between spectra of 10
Myr-old dwarfs, and those of field dwarfs with identical spectral
types (see Fig.~\ref{spectro}). This finding corroborates the age
estimate of HR\,6037\,A and B of a few tens to hundreds Myr (if both
component are coeval).

\begin{table}[!]
\caption{Binary parameters for the two epoch observations.}\label{binary}
\begin{tabular}{llll}     
\hline\hline\noalign{\smallskip}
Date        & Separation & PA    & $\Delta\,K_s$ \\ 
            & (arcsec)      & (deg) & (mag)         \\
\noalign{\smallskip}\hline\noalign{\smallskip}
 2004/06/30  & $6.66 \pm 0.01$   & $293.96\pm0.1$ & $8.2\pm0.1$ \\
 2006/06/09   & $6.67\pm 0.03$   & $293.98\pm0.2$ & $8.7\pm0.2$ \\
\noalign{\smallskip}\hline\noalign{\smallskip}
\end{tabular}
\end{table}

\subsection{HR\,6037\,B physical properties}

The difference in $K_s$ magnitude between HR\,6037\,A\&B in the NACO
images was derived using standard packages for aperture photometry
within IRAF, and is provided in Table~\ref{binary}. Since the 2MASS
$K_{s}$ value of the primary is $5.66\pm0.02$ \citep{2MASS}, we
estimate an average $K_{s}$=$14.1\pm0.3$\,mag for the secondary,
which translates into $M_{K_s}=10.4\pm0.3$\,mag for a distance of
$55\pm2$\,pc.  We have compared this value with evolutionary
tracks by \citet{Baraffe2002}, assuming the age estimate derived from
the UVES spectrum of the primary. According to DUSTY evolutionary
tracks \citep{Baraffe2002}, a $300\pm100$\,Myr object with $M_{K_s}$ of
$10.4\pm0.3$\,mag, corresponds to a $62\pm20$~M$_{Jup}$ BD with
$T_{\rm{eff}} = 2330\pm200$\,K, and log\,$g= 5.1\pm0.2$.

\subsection{HR\,6037 A\&B: main properties of the binary system}

The mass ratio of HR\,6037 A\&B is $q$=0.034. This mass ratio is not
common for binaries with intermediate-mass stars as primaries.  The
projected separation of the binary components, for a distance of
55\,pc, is $\sim$366\,AU.  Using Kepler's third law, we derive an
orbital period of $\sim$5000\,yr.  Even if imaging surveys are
sensitive to these long period, small mass ratio binaries, they are
uncommon \citep[e.g.][]{Kouwenhoven2007,Ehrenreich2010}. Hence, we can
conclude that HR\,6037\,A\&B is a extremely rare binary system.

\section{Conclusions}
We report the detection of a BD companion to the 300\,Myr old star HR\,6037.
Our main results can be summarized as follows:  

\begin{enumerate}
      \item HR\,6037 is a binary system with a  separation of 6\farcs66 and 
      position angle of $293.9\pm0.1$ degrees. Two epoch observations 
      confirm that HR\,6037\,B is a co-moving companion.

      \item Near-IR spectroscopy reveals a spectral type of M9$\pm$1
        for HR\,6037\,B by comparison of the $J$ and $H$ band
        continuum to templates.  The strength of the gravity-sensitive
        features are consistent with a dwarf intermediate between a
        low-gravity young dwarf and high-gravity field dwarf of
        similar spectral type. This result is consistent with the age
        derived for the primary, 300\,Myr, i.e. both objects appear to
        be coeval.

     \item Evolutionary tracks predict a mass of 62$\pm$20\,M$_{Jup}$, 
        an effective temperature of $T_{\rm{eff}} = 2330\pm200$\,K, and a surface gravity of 
        log\,$g= 5.1\pm0.2$.

\end{enumerate}
To our knowledge, HR\,6037\,B is the second BD companion confirmed to be bound to an intermediate-mass star 
by two epoch observations and spectroscopy. Its small mass ratio and long orbital period
makes it a rare and uncommon binary system.

\begin{acknowledgements}
    This research has been funded by Spanish grants
    MEC/ESP2007-65475-C02-02, {\mbox MEC/Consolider-CSD2006-0070, and
      CAM/PRICIT-1496}. We are grateful to K. Allers and E. Rice for
    providing their spectra. NH gratefully acknowledges support from
    the ESO's Visiting Scientist Programme.  GCa thanks S. Villanova
    for helping with the UVES data reduction. This research has made
    use of the SIMBAD database, operated at CDS, Strasbourg, France.
 \end{acknowledgements}.

\bibliographystyle{bibtex/aa}
\bibliography{hip79797}

\end{document}